\documentclass[twocolumn,showpacs,superscriptaddress,amsmath,amssymb]{revtex4}
\usepackage{graphicx}
\usepackage{dcolumn}
\usepackage{bm}
\usepackage{color}

\def\bd{
\begin{document}} \def\ed{\end{document}}
\def\bmp{\begin{minipage}} \def\emp{\end{minipage}}
\def\bcc{\begin{center}} \def\ecc{\end{center}}     \def\npg{\newpage}
\def\beq{\begin{equation}} \def\eeq{\end{equation}} \def\hph{\hphantom}
 \def\r#1{$^{[#1]}$}
\def\n{\noindent} \def\ni{\noindent} \def\pa{\parindent}
\def\hs{\hskip} \def\vs{\vskip} \def\hf{\hfill} \def\ej{\vfill\eject}
\def\cl{\centerline} \def\ob{\obeylines}  \def\ls{\leftskip}
\def\underbar#1{$\setbox0=\hbox{#1} \dp0=1.5pt \mathsurround=0pt
   \underline{\box0}$}   \def\ub{\underbar}    \def\ul{\underline}
\def\f{\left} \def\g{\right} \def\e{{\rm e}} \def\o{\over} \def\d{{\rm d}}
\def\vf{\varphi} \def\pl{\partial} \def\cov{{\rm cov}} \def\ch{{\rm ch}}
\def\la{\langle} \def\ra{\rangle} \def\EE{e$^+$e$^-$} \def\pt{p_{\rm t}}
\def\pti{p_{{\rm t},i}} \def\ptj{p_{{\rm t},j}}
\def\bitz{\begin{itemize}} \def\eitz{\end{itemize}}
\def\btbl{\begin{tabular}} \def\etbl{\end{tabular}}
\def\btbb{\begin{tabbing}} \def\etbb{\end{tabbing}}
\def\beqar{\begin{eqnarray}} \def\eeqar{\end{eqnarray}}
\def\\{\hfill\break} \def\dit{\item{-}} \def\i{\item}
\def\bbb{\begin{thebibliography}{9} \itemsep=-1mm}
\def\ebb{\end{thebibliography}} \def\bb{\bibitem}
\def\bpic{\begin{picture}(260,240)} \def\epic{\end{picture}}
\def\akgt{\cl{\bf ACKNOWLEDGMENTS}}
\def\fgn{\noindent{\bf\large\bf figure captions}}
\def\m1{\langle N_p\rangle} \def\u2{\langle N_{\bar p}\rangle} \def\Nap{N_{\bar
p}}
\def\lan{\langle}
\def\ran{\rangle}
\def\p{\pi}
\def\ifmath#1{\relax\ifmmode #1\else $#1$\fi}%
\def\rc{\ifmath{{\mathrm{c}}}}
\def\cut{\ifmath{{\mathrm{cut}}}}
\def\rF{\ifmath{{\mathrm{F}}}}
\def\rK{\ifmath{{\mathrm{K}}}}
\def\rp{\ifmath{{\mathrm{p}}}}
\def\rt{\ifmath{{\mathrm{t}}}}
\def\LAB{\ifmath{{\mathrm{LAB}}}}
\def\cut{\ifmath{{\mathrm{cut}}}}

\newcommand{\cinst}[2]{$^{\mathrm{#1}}$~#2\par}
\newcommand{\crefi}[1]{$^{\mathrm{#1}}$}
\newcommand{\crefii}[2]{$^{\mathrm{#1,#2}}$}
\newcommand{\crefiii}[3]{$^{\mathrm{#1,#2,#3}}$}
\newcommand{\HRule}{\rule{0.5\linewidth}{0.5mm}}

\bd
\title{Statistical and dynamical fluctuations in the ratios of higher net-proton cumulants\\ 
in relativistic heavy ion collisions}


\author{Lizhu Chen} 
\affiliation{Institute of Particle Physics, Hua-Zhong Normal
University, Wuhan 430079, China}\affiliation{Brookhaven National
Laboratory, Upton, NY 11973, U.S.A.}
\author{Xue Pan} 
\affiliation{Institute of Particle Physics, Hua-Zhong Normal
University, Wuhan 430079, China}
\author{Fengbo Xiong} 
\affiliation{Institute of Particle Physics, Hua-Zhong Normal
University, Wuhan 430079, China}
\author{Lin Li} 
\affiliation{Institute of Particle Physics, Hua-Zhong Normal
University, Wuhan 430079, China}
\author{Na Li} 
\affiliation{Institute of Particle Physics, Hua-Zhong Normal
University, Wuhan 430079, China}
\author{Zhiming Li} 
\affiliation{Institute of Particle Physics, Hua-Zhong Normal
University, Wuhan 430079, China}
\author{Gang Wang}
\affiliation{Department of Physics and Astronomy, University of California, Los Angeles, CA 90095, U.S.A.}
\author{Yuanfang Wu} 
\affiliation{Institute of Particle Physics,
Hua-Zhong Normal University, Wuhan 430079,
China}\affiliation{Brookhaven National Laboratory, Upton, NY 11973,
U.S.A.}\affiliation{Key Laboratory of Quark $\&$ Lepton Physics
(Huazhong Normal University), Ministry of Education, China }

\begin{abstract}
With the help of transport and statistical models, we find that
the ratios of higher net-proton cumulants measured at RHIC 
are dominated by the statistical fluctuations.
Future measurements should focus on the dynamical fluctuations,
which are relevant to the underlying mechanisms of particle production, the critical phenomena in particular.
We also demonstrate that a proton-antiproton correlation 
directly show if protons and antiprotons are emitted independently. 
\end{abstract}


\maketitle
\section{Introduction}
One of the main goals of relativistic heavy ion collisions is to locate 
the critical point on the QCD phase diagram, spanned by the
temperature ($T$) and the baryon chemical potential ($\mu_B$).
At the critical point,  the correlation length ($\xi$) goes to infinity and the long range correlations become dominant. 
So the fluctuations of final state particles are expected to be largely enhanced,
if the $(T, \mu_B)$ trajectory of the collision system is close to the critical point.
The $\xi$-related observables are therefore of great interest in heavy ion collisions~\cite{corr-fluc}.

The thermodynamic quantities, such as order parameter, specific heat capacity and susceptibility ($\chi$), diverge with the correlation length at the critical point.
The $i$th net-baryon cumulant is recently shown to be directly related to the $i$th susceptibility ($\chi_i$) of the formed system~\cite{antoniou,stephanov, koch},
\begin{equation}\label{susceptibility-1}
  \langle \delta N^i\rangle=VT\chi_i,
\end{equation}
\noindent where $N$ is the net-baryon number,
$ \langle \delta N^i\rangle =\langle (N-\langle N\rangle)^i \rangle$ is the $i$th net-baryon cumulant, and $V$ is the volume.
The third and fourth cumulants,
\begin{equation}\label{k34}
  K_3=\langle\delta N^3\rangle, \ \
  K_4=\langle\delta N^4\rangle-3\langle\delta N^2\rangle^2,
\end{equation}
\noindent are argued to be more sensitive to the correlation length 
as they are proportional to $\xi^{4.5}$ and $\xi^7$, respectively~\cite{stephanov,koch,rajargopal,akasawa}.
Experimentally, the proton number is a good approximation of the baryon number~\cite{stephanov},
and the properly normalized ratios, net-proton Skewness and Kurtosis,
\beqar\label{kurtosis-ex}
 S=K_3/K_2^{3/2}=\frac{\langle\delta N^3\rangle}{\langle\delta
  N^2\rangle^{3/2}},\nonumber\\
 K=K_4/K_2^2=\frac{\langle\delta N^4\rangle}{\langle\delta
  N^2\rangle^2}-3,
\eeqar \noindent are preferred~\cite{star-prl}, 
measuring the symmetry and sharpness of the net-proton distribution, respectively.

The STAR measurements~\cite{star-prl} show that both net-proton Skewness and
Kurtosis decrease with increasing number of participants (centrality),
which could be explained by the central limit theorem (CLT).
On the other hand, various model calculations~\cite{luoxf} reproduce the experimental results surprisingly well,
raising the suspicion that Skewness and Kurtosis
are insensitive to the mechanisms of particle production implemented in different models.
Recently, Karsch and Redlich~\cite{karsch} have derived
simple relations between the cumulant ratios and the thermal parameters
($T$ and $\mu_B$ at the chemical freeze-out~\cite{parameters}),
based on the hadron resonance gas (HRG) model~\cite{HRG},
with the system well thermalized and without phase transition. 
They have shown that the HRG model results are well
consistent with the STAR data at different collision energies~\cite{karsch}.

However, before trying to understand the physics behind the measured Skewness and Kurtosis, 
we need to take into account and properly remove the contributions from the statistical fluctuations~\cite{bialas,rajargopal} 
and all non-thermal sources (minijets, resonance decay, initial size fluctuation, etc.)~\cite{gupta, Giorgio}.
In this paper, we will focus on the elimination of the statistical fluctuations. 
We first estimate the statistical fluctuations in the cumulant ratios,
which turn out to dominate the behavior of net-proton Kurtosis at RHIC energies. 
Then we propose the dynamical ratios of higher net-proton cumulants in Section III
and the correlation between proton and antiproton in Section IV. 
The centrality dependence of the dynamical ratios and the correlations
from two versions of AMPT~\cite{ampt}, UrQMD~\cite{urqmd},  and
Therminator~\cite{therminator} are presented and discussed. 
Finally, the summary and conclusions are given in Section V.

\section{Statistical fluctuations in the ratios of higher net-proton cumulants }

The statistical fluctuation comes from the finite number of particles,
usually obeying a Poisson distribution~\cite{bialas,claude,rajargopal}. 
If we have two independent Poisson distributions for protons and antiprotons
with means $\m1$ and $\u2$, respectively, 
the net-proton number ($N$) follows a Skellam (SK) distribution~\cite{luoxf,skellam},
\begin{align}
&f(N;\m1,\u2)\nonumber\\&=e^{-(\m1+\u2)}(\m1/\u2)^{\frac{N}{2}}I_{|N|}\left(2\sqrt{\m1\u2}\right),
\end{align}
\noindent where $I_{|N|}(2\sqrt{\m1\u2})$ is the modified Bessel function of the first kind.
Then the statistical fluctuations of the ratios of higher net-proton
cumulants can be directly deduced from the Skellam distribution,
\beqar\label{st-KS}
S_{\rm stat}&=&\frac{\m1-\u2}{[\m1+\u2]^{3/2}},\nonumber \\
K_{\rm stat}&=&\frac{1}{\m1+\u2}, \nonumber\\
R_{2,1,{\rm stat}}&=&\frac{\m1+\u2}{\m1-\u2},\nonumber\\
R_{3,2,{\rm stat}}&=&\frac{\m1-\u2}{\m1+\u2},\nonumber\\
R_{4,2,{\rm stat}}&=& 1,\eeqar \noindent  
where $R_{i,j} = \chi_i / \chi_j=K_i/K_j$.
The HRG model results~\cite{karsch} show that $R_{4,2}$ is unity,
and now we find that it is completely statistical.
The other ratios are determined only by the mean numbers of protons and
antiprotons, which usually increase with the incident energy
and centrality. So the statistical fluctuations of Skewness and Kurtosis 
decrease with the incident energy and centrality.

In Fig.~1(a), the model results of net-proton Kurtosis and its statistical fluctuations (SK)
are shown from APMT default, AMPT with string melting~\cite{ampt} and Therminator~\cite{therminator}. Where the transverse momentum, $\pt$,  and rapidity cuts are respectively $0.4<\pt<0.8$ GeV/c, $ |y|<0.5 $,  the same as they are given at RHIC/STAR published paper~\cite{star-prl}. In both transport (AMPT) and statistical (Therminator) models, 
the net-proton Kurtosis results (solid points) are very close to the corresponding 
statistical fluctuations (open points).
The STAR measurements of net-proton Kurtosis (stars)~\cite{star-prl} follow the same trend
as a function of the number of participants ($N_{\rm part}$).
So the statistical fluctuations dominate the behavior of net-proton Kurtosis at RHIC,
which is why the results from various models closely resemble the experimental data.

\begin{figure}
\includegraphics[width=3.3in]{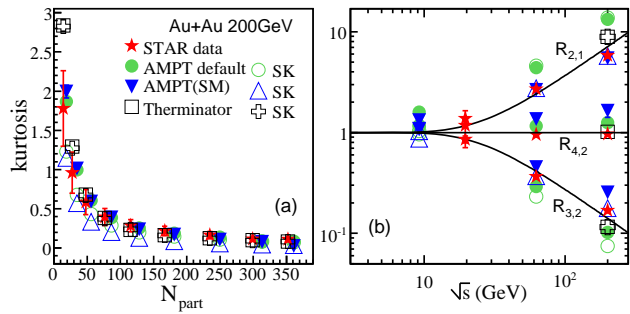}
\caption{\label{Fig. 1}(Color online) (a) Centrality dependence of
net-proton Kurtosis at 200 GeV, and (b) Energy
dependence of $R_{2,1}$, $R_{4,2}$ and $R_{3,2}$, for Au + Au collisions. 
The results are obtained from RHIC/STAR data~\cite{star-prl}, HRG~\cite{karsch}, 
two versions of AMPT~\cite{ampt} and Therminator~\cite{therminator} models, 
and the corresponding statistical fluctuations are estimated with the Skellam distribution (SK) from 
Eq.~(\ref{st-KS}), respectively.}
\end{figure}

In Fig.~1(b), we show the incident-energy dependence of the net-proton ratios,
$R_{2,1}$, $R_{4,2}$, and $R_{3,2}$, obtained from RHIC/STAR
data, HRG, AMPT default,  AMPT with string melting and Therminator,
together with the corresponding statistical fluctuations (SK) of the last three cases determined 
by Eq.~(5). At 200 GeV,  the ratios from Therminator (squares) coincide with the corresponding statistical fluctuations (crosses),
because Therminator is a statistical model with only constraints on kinetics.
The HRG curves~\cite{karsch} are given by $R_{2,1}=1/\tanh(\mu_B/T)$ and 
$R_{3,2}= \tanh(\mu_B/T)$. Similarly, both Therminator and HRG models have protons and antiprotons completely independently emitted with Poisson distributions,
so their results are close to the data from RHIC/STAR, where the statistical fluctuations dominate.
The ratios from two versions of AMPT (solid circles and triangles) slightly deviate from the corresponding statistical fluctuations (open circles and rhombi), and from the HRG lines,
due to the more complicated particle production mechanisms in AMPT.  

We have demonstrated that the influence of statistical fluctuations is far from being
negligible in the ratios of higher net-proton cumulants at RHIC energies.
To investigate the underlying physics, we have to first remove the statistical fluctuations.

\section{Dynamical ratios of higher net-proton cumulants }

There have been long efforts in eliminating the statistical fluctuations in
elementary collisions~\cite{bialas, kittel}. For a single particle distribution,
the factorial moments are used to remove the statistical fluctuations~\cite{antoniou,bialas}.
But this method can not be directly generalized to 
the distribution of the difference between two Poisson variables.

From previous discussions, the statistical fluctuations in the ratios of net-proton cumulants
are directly obtainable. 
The dynamical ratios of net-proton cumulants can be simply defined as a
deviation of the ratios from the statistical fluctuations~\cite{claude},
e.g., \beqar\label{dyn-KS} 
K_{\rm dyn}&=&K-K_{\rm stat},
\eeqar 
\noindent and so on, where the statistical parts are given by Eq.~(\ref{st-KS}).

\begin{figure}
\includegraphics[width=3.4in]{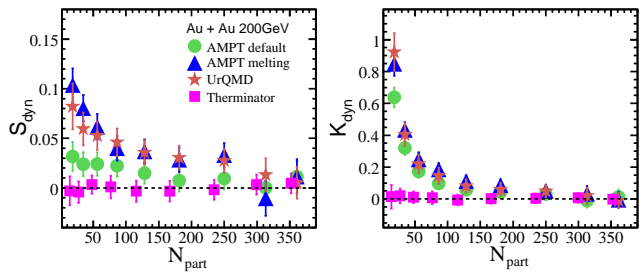}
\caption{\label{Fig. 2}(Color online) Centrality dependence of
dynamical Skewness (left) and Kurtosis (right) for Au + Au
collisions at 200 GeV, given by transport models (AMPT and UrQMD)
and a statistical model (Therminator).}
\end{figure}

The centrality dependence of dynamical net-proton Skewness and Kurtosis 
are shown in Fig.~2(a) and (b), respectively,
from AMPT default, AMPT with string melting, UrQMD and Therminator models for Au + Au collisions at 200 GeV. 
Both dynamical Skewness and Kurtosis from Therminator are zero at all centralities,
illustrating that the symmetry and sharpness of the net-proton distribution in the model 
both follow the Skellam distribution.
For the transport models (AMPT and UrQMD), both dynamical Skewness and Kurtosis are larger than zero in peripheral collisions, 
and approach zero in central collisions. Compared with the Skellam distribution, 
the positive dynamical Kurtosis and Skewness implies that the net-proton distribution has a sharper peak and a longer tail at the large net-proton side, respectively. 
These deviations are caused by non-thermal sources implemented in transport models.


To study how the results change with the incident energy, 
the centrality dependence of dynamical net-proton Skewness and Kurtosis from AMPT default are shown in Fig.~3
for Au+Au collisions at 3 incident energies. 
When the incident energy changes from 200 GeV to 39 GeV, both dynamical Skewness and Kurtosis remain positive.
Dynamical Skewness shows a significant dependence on the incident energy, especially in peripheral collisions,
and dynamical Kurtosis is almost independent of the incident energy.

\begin{figure}
\includegraphics[width=3.4in]{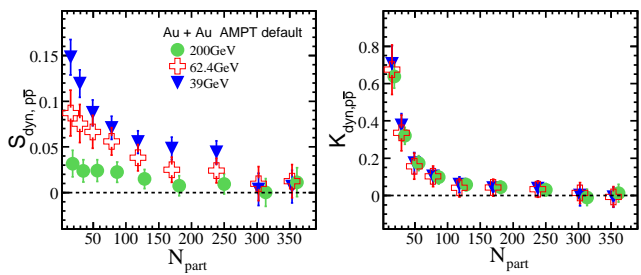}
\caption{\label{Fig. 3}(Color online)  Centrality dependence of
dynamical net-proton Skewness (left) and Kurtosis (right) for Au + Au collisions at 3 incident energies,
obtained from AMPT default.}
\end{figure}

The behavior of dynamical ratios of higher cumulants are called for at the RHIC beam energy scan. 
If the deviation from the statistical fluctuations is zero, 
protons and antiprotons are emitted independently as the statistical models assume. 
Otherwise if the non-zero deviation remains the same sign for different incident energies, 
like what the transport model shows in Fig.~3, then there is no critical related phenomena. 
However, if the deviation changes dramatically with the variation of the incident energy, 
e.g. showing the sign changes at the third and fourth cumulants, 
where the symmetry and sharpness of the net-proton distribution deviate from 
the corresponding statistical fluctuations in opposite directions~\cite{akasawa, Liuyx, fs3}, 
it may reveal the critical incident energy nearby~\cite{fs3}.

\section{Correlations between proton and antiproton}

To see if protons and antiprotons are emitted independently,
we could also directly measure the correlation
between them, \beqar{\label{corr}}
C(N_p,N_{\bar p})=\frac{\la N_pN_{\bar p}\ra}{\la N_p\ra\la N_{\bar
p}\ra}-1.\eeqar 
\noindent The correlation will be zero if protons and antiprotons are independent.

The centrality dependence of the correlations from two versions of
AMPT, UrQMD and Therminator models for Au + Au collisions at 200 GeV are presented in Fig.~4. 
The correlation is zero at all centralities in Therminator,
another illustration of the model's assumption.
In transport models, the correlation decreases with centrality,
following a similar trend as dynamical Skewness or Kurtosis. The correlation is positive, 
indicating that protons and antiprotons are not emitted independently.
This leads to the difference between the net-proton distribution in transport models and the pure statistical Skellam distribution.

\begin{figure}
\includegraphics[width=2.5in]{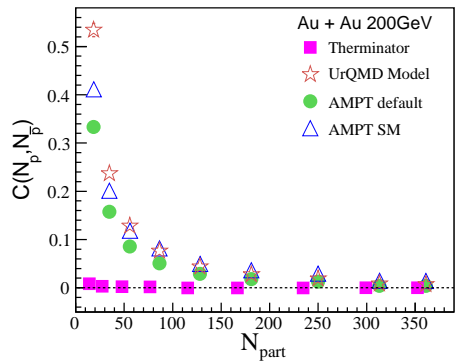}
\caption{\label{Fig. 4}(Color online)  Centrality dependence of
proton antiproton correlations for Au + Au collisions at 200 GeV,
given by transport models (AMPT and UrQMD)
and a statistical model (Therminator).}
\end{figure}


\section{Summary}

We have demonstrated that the statistical fluctuations dominate the behavior of the ratios of 
higher net-proton cumulants measured at RHIC,
and this explains why the results from various models are consistent with the experimental data.
We argue that before trying to understand the underlying physics the
statistical fluctuations should be taken into account. 

To study the particle production mechanism, the dynamical ratios of higher net-proton cumulants
and the correlations between proton and antiproton have been proposed and discussed. 
It is shown that the dynamical ratios and the correlations are similarly zero at all centralities in a statistical model, and positive in transport models. This indicates that protons 
and antiprotons are not emitted independently in transport models. 

The behaviors of the dynamical ratios of higher net-proton cumulants, as well as that of the proton-antiproton correlation,
are more relevant to the location of the critical point, and the corresponding measurements during RHIC beam energy scan
will shed light on the study of the QCD phase transition.

\vskip 0.5cm

We are grateful for stimulating discussions with Dr. Nu Xu, Xiaofeng Luo, Dr. Fuqiang Wang and Dr. Zhangbu Xu. The first and last authors are grateful for the hospitality of BNL STAR group. 
This work is supported in part by the NSFC of China with project No. 10835005, 11005046, MOE of China with project No. IRT0624, No. B08033 and a grant from U.S. Department of Energy, Office of Nuclear Physics.

\ed